\documentclass[preprint,showpacs,preprintnumbers,nofootinbib,superscriptaddress]{revtex4}
\usepackage{graphicx}
\usepackage{epsfig}
\usepackage{dcolumn}
\usepackage{bm}

\def\gsim{\lower0.5ex\hbox{$\:\buildrel >\over\sim\:$}}
\def\lsim{\lower0.5ex\hbox{$\:\buildrel <\over\sim\:$}}

\newcommand{\bea}{\begin{eqnarray}}
\newcommand{\eea}{\end{eqnarray}}

\def\mpT{p_T \hspace{-.9em}/\;\:}

\begin{document}

\preprint{HIP-2005-43/TH}

\title{Signals of sneutrino-antisneutrino mixing in an $e^- \gamma$  
collider in anomaly-mediated supersymmetry breaking}

\author{Tuomas Honkavaara}
\email{tuomas.honkavaara@helsinki.fi}
\affiliation{High Energy Physics Division, Department of Physical 
Sciences, P.O. Box 64, FIN-00014 University of Helsinki, Finland}
\affiliation{Helsinki Institute of Physics, P.O. Box 64, FIN-00014  
University of Helsinki, Finland}
\author{Katri Huitu}
\email{Katri.Huitu@helsinki.fi}
\affiliation{High Energy Physics Division, Department of Physical 
Sciences, P.O. Box 64, FIN-00014 University of Helsinki, Finland}
\affiliation{Helsinki Institute of Physics, P.O. Box 64, FIN-00014 
University of Helsinki, Finland}
\author{Sourov Roy}
\email{roy@cc.helsinki.fi}
\affiliation{Helsinki Institute of Physics, P.O. Box 64, FIN-00014 
University of Helsinki, Finland}

\received{\today}


\begin{abstract} \vspace*{10pt}
Sneutrino-antisneutrino mixing occurs in a supersymmetric model 
where neutrinos have nonzero Majorana masses. This can lead to 
the sneutrino decaying into a final state with a ``wrong-sign charged lepton". 
Hence, in an $e^- \gamma$ collider, the signal of the associated production 
of an electron-sneutrino and the lighter chargino and their subsequent decays 
can be $e^- \gamma \rightarrow e^+ {\tilde \tau}_1^- {\tilde \tau}_1^- + \mpT$ 
where the ${\tilde \tau}_1$s are long-lived and can produce heavily ionizing 
charged tracks. This signal is free of any Standard Model background and the 
supersymmetric backgrounds are small. Such a signal can be experimentally observable 
under certain conditions which are possible to obtain in an anomaly-mediated 
supersymmetry breaking scenario. Information on a particular combination of the 
neutrino masses and mixing angles can also be extracted through the observation 
of this signal. Possible modifications in the signal event and the accompanying
Standard Model background have been discussed when the ${\tilde \tau}_1$s decay 
promptly. 

\end{abstract}

\maketitle

\section{Introduction}
There has been a tremendous experimental progress in neutrino physics in 
recent years, and the present data from the solar and  
atmospheric neutrino experiments contain compelling evidence that 
neutrinos have tiny masses \cite{altarelli-kayser}. It is 
widely believed that the lepton number ($L$) may be violated in 
nature and the neutrinos are Majorana particles. In this case, the  
smallness of the neutrino masses can be explained by the seesaw mechanism 
or by dimension-five non-renormalizable operators with a generic structure. In 
the context of supersymmetric theories, such $\Delta L$ = 2 
Majorana neutrino mass terms can induce mixing between the sneutrino and 
the antisneutrino and a mass splitting ($\Delta m_{\tilde \nu}$) between 
the physical states \cite{hirschetal,grossman-haber1,grossman-haber2,chun,
davidson-king}. The effect of this mass splitting is to induce sneutrino-antisneutrino 
oscillations, and the lepton number can be tagged in sneutrino decays by 
the charge of the final state lepton. This situation is similar to the 
flavour oscillation in the $B^0$--$\bar B^0$ system \cite{bsystem}. 
Suppose the physical sneutrino states are denoted by $|{\tilde \nu}_1 \rangle$ and 
$|{\tilde \nu}_2\rangle$. An initially (at $t=0$) produced pure $|\tilde \nu 
\rangle$ state is related to the mass eigenstates as 
\bea
|\tilde \nu \rangle = \frac{1}{\sqrt2}[|{\tilde \nu}_1 \rangle + i|{\tilde 
\nu}_2 \rangle].
\eea
The state at time $t$ is 
\bea
|{\tilde \nu}(t) \rangle = \frac {1} {\sqrt{2}}[e^{-i(m_1-i\Gamma_{\tilde 
\nu}/2)t} |{\tilde \nu}_1 \rangle + ie^{-i(m_2-i\Gamma_{\tilde \nu}/2)t}
|{\tilde \nu}_2 \rangle],
\eea
where the difference between the total decay widths of the two mass 
eigenstates has been neglected, and the total decay width is set to be 
equal to $\Gamma_{\tilde \nu}$. Since the sneutrinos decay, the probability 
of finding a ``wrong-sign charged lepton" in the decay of a sneutrino should be the 
time-integrated one and is given by
\bea
P(\tilde \nu \rightarrow \ell^+) = {\frac {x^2_{\tilde \nu}} 
{2(1+x^2_{\tilde \nu})}} \times B({\tilde \nu}^* \rightarrow \ell^+),
\label{eqn-prob}
\eea
where the quantity $x_{\tilde \nu}$ is defined as 
\bea
x_{\tilde \nu} \equiv \frac {\Delta m_{\tilde \nu}} {\Gamma_{\tilde \nu}},
\label{xnu}
\eea  
and $B(\tilde \nu^* \rightarrow \ell^+)$ is the branching fraction for 
$\tilde \nu^* \rightarrow \ell^+$. Here, we assume that sneutrino flavour 
oscillation is absent and the lepton flavour is conserved in the decay of 
antisneutrino/sneutrino. 
If $x_{\tilde \nu} \sim 1$ and if the branching ratio of the antisneutrino into the
corresponding charged lepton final state is also significant, then one can have a 
measurable ``wrong-sign charged lepton" signal from the single production of a sneutrino 
in colliders. In a similar way, lepton flavour oscillation has been discussed in 
Ref. \cite{leptonflavor}. 

It is evident from the above discussion that the probability of the 
sneutrino-antisneutrino oscillation depends crucially on $\Delta m_{\tilde 
\nu}$ and $\Gamma_{\tilde \nu}$. Taking into account the radiative 
corrections to the Majorana neutrino mass $m_\nu$ induced by $\Delta 
m_{\tilde \nu}$, one faces the bound \cite{grossman-haber1} $\Delta 
m_{\tilde \nu}/m_\nu \lsim \mathcal O (4\pi/\alpha)$. If we consider 
$m_\nu$ to be $\sim$ 0.1 eV, then $\Delta m_{\tilde \nu} \lsim 0.1$ keV. 
Thus, in order to get $x_{\tilde \nu} \sim 1$, one also needs the 
sneutrino decay width $\Gamma_{\tilde \nu}$ to be $\sim 0.1$ keV or so. In 
other words, this small decay width means that the sneutrino should 
have enough time to oscillate before it decays. However, such a small  
decay width is difficult to obtain in most of the scenarios widely  
discussed in the literature with the lightest neutralino ($\tilde 
\chi^0_1$) being the lightest supersymmetric particle (LSP). In 
this case, the two-body decay channels $\tilde \nu \rightarrow \nu 
\tilde\chi^0$ and/or $\tilde \nu \rightarrow \ell^- \tilde \chi^+$ 
involving 
the neutralinos ($\tilde \chi^0$) and the charginos ($\tilde \chi^+$) will 
open up. In order to have a decay width $\Gamma_{\tilde \nu} \lsim {\cal 
O}(1)$ keV, these two-body decay modes should be forbidden so that the 
three-body decay modes $\tilde \nu \rightarrow \ell^- {\tilde \tau}_1^+ 
\nu_\tau$ and $\tilde \nu \rightarrow \nu {\tilde \tau}_1^\pm \tau^\mp$ 
are the available ones. In addition, one should get a reasonable branching fraction 
for the $\ell^- {\tilde \tau}_1^+ \nu_\tau$ final state in order to get
the wrong-sign charged lepton signal. It has been pointed out in 
Ref. \cite{grossman-haber1} that, in order to achieve these 
requirements, one should have a spectrum 
\bea
m_{{\tilde \tau}_1} < m_{\tilde \nu} < m_{\tilde \chi^0_1}, m_{\tilde 
\chi^\pm_1}, 
\label{spectrum}
\eea   
where the lighter stau (${\tilde \tau}_1$) is the LSP. However, having 
${\tilde \tau}_1$ as a stable charged particle is strongly disfavoured by 
astrophysical grounds \cite{staulsp}. This could be avoided, for example, 
by assuming a very small $R$-parity-violating coupling which induces the 
decay ${\tilde \tau}_1 \rightarrow \ell \nu$ but still allows 
${\tilde\tau}_1$ to have a large enough decay length to produce a heavily 
ionizing charged track inside the detector. As we will discuss later on, 
the spectrum (\ref{spectrum}) can be obtained in some part of the parameter 
space in the context of anomaly-mediated supersymmetry breaking (AMSB) 
with $\Delta m_{\tilde \nu} \lsim \mathcal O (4\pi m_\nu/\alpha)$. Hence, 
AMSB seems to have a very good potential to produce signals of 
sneutrino-antisneutrino oscillation which can be tested in 
colliders. 

Like-sign dilepton signals from sneutrino-antisneutrino mixing with or 
without $R$-parity have been discussed in the context of an $e^+e^-$ 
linear collider and hadron colliders 
\cite{grossman-haber1,like-sign-rp,like-sign-norp}. In the context of 
$R$-parity-conserving supersymmetry, like-sign dilepton signal has also 
been calculated in an $e^-\gamma$ collider \cite{like-sign-rp}. Some other 
phenomenological implications of sneutrino-antisneutrino mass splitting 
have also been discussed in Refs. \cite{majoranasnu,shaouly} for various 
present and future colliders. In this paper, we consider the signal of 
sneutrino-antisneutrino oscillation via the observation of a ``wrong-sign 
charged lepton" in the context of an $e^-\gamma$ collider. In particular, 
we look at the process $e^-\gamma \rightarrow {\tilde \nu}_e 
\tilde \chi^-_1$ which will eventually lead to the final state $e^+ {\tilde 
\tau}_1^- {\tilde \tau}_1^- + \mpT$ if the initial ${\tilde \nu}_e$ 
oscillates into a ${\tilde \nu}_e^*$. The long-lived staus will then 
produce two heavily ionizing charged tracks, making this signal rather 
unique in the context of an $e^- \gamma$ collider. As discussed above, 
AMSB can give an observable rate for such a signal. In section II of our 
paper, we will briefly discuss the basic features of the AMSB scenario 
which we consider, and calculate the sneutrino-antisneutrino mass 
splitting. In this respect, we will broadly follow the philosophy of Ref. 
\cite{like-sign-norp}. We will also calculate the total decay width of the 
sneutrino/antisneutrino and the probability of the sneutrino/antisneutrino 
decaying to a final state with a ``wrong-sign charged lepton" given in Eq. 
(\ref{eqn-prob}). Section III discusses in brief the properties of an 
$e^-\gamma$ collider and the photon spectrum. In Section IV, numerical 
results of our calculation of the signal and the background in the available 
region of the parameter space are discussed, and we finally conclude in Section V.

\section{Anomaly mediation and sneutrino-antisneutrino mixing}

In anomaly-mediated supersymmetry breaking, one assumes that the hidden 
sector and the observable sector superfields are not directly 
connected, e.g. they are
localized on two 
different parallel 3-branes in higher dimensions. These branes are 
separated by a distance of the order of the compactification radius $r_c$ along 
the extra dimension. SUSY breaking is communicated by the super-Weyl anomaly through 
the Weyl compensator superfield $\Phi_0$ of the supergravity multiplet 
\cite{amsb1}: 
\bea
\Phi_0 = 1 + \theta^2 F_{\Phi_0},
\eea 
where $F_{\Phi_0}$ is $\mathcal O(m_{3/2})$, the gravitino mass. 

In its simplest form, the AMSB scenario predicts tachyonic sleptons and 
should hence be modified. 
Theoretical motivations for suitable modifications have been discussed
e.g. in \cite{amsbmod}.  
Here, we adopt the minimal model in which one assumes that a 
universal term $m^2_0$ is added to all the soft scalar squared masses at 
the grand unified theory (GUT) scale $M_{GUT} \approx 2 \times 10^{16}$ 
GeV. The expression for the scalar masses is given by
\bea
m^2_i = m^2_0 - \frac {1} {4} {\frac {d\gamma_i}{d \ln Q}}{|F_{\Phi_0}|}^2,
\label{eqn-scalar}
\eea 
where $\gamma_i = d \ln Z_i/d \ln Q$ is the anomalous dimension. The gaugino masses 
are given by
\bea
M_i = {\frac {{b_i} {g^2_i}} {16\pi^2}}F_{\Phi_0},
\label{eqn-gaugino}
\eea
where $b_i$ = (33/5,1,$-3$) are the one-loop beta function coefficients 
for the ${\mathrm U(1)}$, ${\mathrm {SU}}(2)$ and ${\mathrm {SU}}(3)$ gauge couplings,
respectively. For the trilinear soft SUSY breaking parameters, one has
\bea
A_{ijk} = {\frac {1}{2}} (\gamma_i + \gamma_j + \gamma_k)F_{\Phi_0}.
\label{eqn-trilinear}
\eea 

The minimal AMSB (mAMSB) model is described by the following parameters$\colon$ the
gravitino mass $m_{3/2}$, the common scalar mass parameter $m_0$, the ratio of
Higgs vacuum expectation values $\tan\beta$ and the sign of the higgsino mass
parameter sign($\mu$). The characteristic signatures of the mAMSB model with 
a wino LSP have been studied in the context of hadron colliders 
\cite{amsb-hadron1,amsb-hadron2}, as well as for high energy 
linear colliders \cite{amsb-linearee,amsb-lineareg-gg}. A brief review on the 
signals of the mAMSB model in linear colliders can be found in Ref. \cite{reviewamsb}. 
In this work, we will concentrate on an $e^- \gamma$ collider and discuss the 
signatures of an AMSB model which can accommodate a small Majorana mass for the 
neutrino and consequently generate a $\Delta L = 2$ sneutrino mass splitting.

In order to generate small neutrino masses in this scenario, one should 
include the dimension-5 operators in the effective superpotential at the 
weak scale and also the associated soft SUSY breaking interactions 
\cite{like-sign-norp}. The high energy SUSY preserving dynamics, which generates 
the small neutrino masses, can be the exchange of a heavy right-handed neutrino 
with mass $M$ or the exchange of a heavy triplet Higgs boson. Here, we assume that 
the scale $M$ is far above the weak scale. 
The relevant part of the superpotential and the soft 
SUSY breaking interactions are given by \cite{like-sign-norp}
\bea
\Delta W_{\rm eff} = \frac {\Phi_0}{M}\lambda_{ij}(L_iH_2)(L_jH_2),
\label{eqndelw}
\eea 
\bea
\Delta{\cal L}_{\rm soft} = \frac {C_{ij}\lambda_{ij}}{M}({\tilde \ell}_i h_2) 
({\tilde \ell}_j h_2),
\label{eqndelsoft}
\eea
where $H_2$ is the Higgs doublet superfield giving masses to the 
up-type quarks and $L_i$ are the lepton doublet superfields. The scalar 
components of $L_i$ and $H_2$ are denoted by ${\tilde \ell}_i$ and $h_2$, 
respectively, and $C_{ij} \approx F_{\Phi_0}$. $\lambda$ is a matrix in 
flavour space.

Once the electroweak symmetry is broken, a neutrino mass matrix is 
generated from Eq. (\ref{eqndelw}) and is given by
\bea
(m_\nu)_{ij} = {\frac {2}{M}} \lambda_{ij}{\langle h_2 \rangle}^2.  
\eea
The operator in Eq. (\ref{eqndelsoft}) gives rise to the $\Delta L = 2$ 
sneutrino mass-squared matrix given by
\bea
\frac{1}{2} {\left(\frac{1}{2}\Delta m^2_{\tilde \nu}\right)}_{ij} 
{\tilde \nu}_i {\tilde \nu}_j + \mathrm{H.c.},
\eea
where ${(\Delta m^2_{\tilde \nu})}_{ij} = 
2 (C_{ij}-2\mu{\cot\beta})(m_{\nu})_{ij}$. In addition, sneutrinos have also the 
usual ``Dirac" masses which are written as
\bea
(m^2_{\tilde \nu})_{ij}{\tilde \nu}^*_i {\tilde \nu}_j,
\eea 
where $(m^2_{\tilde \nu})_{ij} \approx m^2_{\tilde \nu} \delta_{ij}$ with 
$m^2_{\tilde \nu} = \frac {1}{2} M^2_Z {\rm cos}2\beta + m^2_{\tilde 
\ell}$, and the slepton doublet mass-squared matrix is assumed to be of 
the form $(m^2_{\tilde \ell})_{ij} \approx m^2_{\tilde \ell} \delta_{ij}$. 

In the AMSB scenario, $C_{ij} \approx F_{\Phi_0}$ and ${F_{\Phi_0}}/{m_{\tilde 
\nu}} = \mathcal O(4\pi/\alpha)$.  Using the relation $\Delta m^2_{\tilde \nu} = 
2 m_{\tilde \nu} \Delta m_{\tilde \nu}$, we can write 
\bea
(\Delta m_{\tilde \nu})_{ij} \approx {\frac {F_{\Phi_0}} {m_{\tilde \nu}}} 
(m_\nu)_{ij} = \mathcal O (4\pi(m_\nu)_{ij}/\alpha).
\label{eqn-deltamsnu}
\eea
Since we want to produce an electron-sneutrino, the relevant 
sneutrino-antisneutrino mass splitting in our case is given by $(\Delta 
m_{\tilde \nu})_{ee} = \frac{4\pi}{\alpha}(m_\nu)_{ee}$, where we have 
neglected the effects suppressed by ${\Delta  m_{\tilde \nu}}/{\delta  
m_{\tilde \nu}}$. Here, $\delta m_{\tilde\nu}$ represents the deviation 
from the exact degeneracy of the $\Delta L$ = 0 sneutrino masses, and 
${\delta  m_{\tilde \nu}} \gg {\Delta  m_{\tilde \nu}}$. Thus, for a given 
neutrino mass, the AMSB model predicts a larger sneutrino-antisneutrino 
mass splitting compared to the models where ${C_{ij}}/{m_{\tilde \nu}}$ is 
$\mathcal O(1)$. As mentioned in the Introduction, it is also possible to 
have the mass spectrum (\ref{spectrum}) in a significant portion of the allowed 
region of the parameter space of the minimal AMSB model, which can lead to a small 
decay width of the sneutrino ($\Gamma_{\tilde \nu} \lsim$ 1 keV). These two 
features make the minimal AMSB model a potential candidate to produce a 
sizeable ``wrong-sign charged lepton" signal in an $e^-\gamma$ collider. 
In section IV of our paper, we will show the allowed region of the 
parameter space where an appreciable number of signal events can be seen. 

As we know, the neutrino oscillation experiments determine only the 
mass-squared differences, but not the absolute scale of the neutrino masses. 
Information on the sum of  the neutrino masses can be obtained from the 
galaxy power spectrum combined with the measurements of the cosmic microwave 
background anisotropies \cite{altarelli-kayser,bilenky}. Several recent 
analyses of cosmological data \cite{cosmo-neut}, which are using results 
of different measurements, give an upper limit in the range $\Sigma_i |m_i| 
\leq$ (0.4--1.7) eV (at 95$\%$ C.L.). However, if we consider only the lower end 
of this limit, then we have $m_\nu \lsim 0.14$ eV for three degenerate neutrinos 
of mass $m_\nu$. On the other hand, neutrinoless double beta decay provides direct 
information on the absolute scale of the neutrino masses. The neutrinoless double 
beta decay is also important due to the fact that it is related to the lepton number 
violating Majorana mass of the neutrino. The present limit from the neutrinoless 
double beta decay is $|(m_\nu)_{ee}| \leq$ 0.2 eV \cite{altarelli-kayser,bilenky}, 
where $(m_\nu)_{ee} = \Sigma U^2_{ei}m_i$ in terms of the mixing matrix ($U_{ei}$) 
and the mass eigenvalues ($m_i$). Recently, some evidence for the neutrinoless double 
beta decay has been reported \cite{neutrinoless}. If this result were 
confirmed, it would favour the degenerate neutrino scenario. From Eq. 
(\ref{eqn-deltamsnu}) and the probability (\ref{eqn-prob}), one can see 
that the ``wrong-sign charged lepton" signal depends on the neutrino mass 
matrix elements $(m_\nu)_{ee}$. Thus, we see that one can get information 
on $(m_\nu)_{ee}$ also from sneutrino-antisneutrino mixing 
\cite{sneu-double,like-sign-norp}. It is important to note here that the 
one-loop contribution to the neutrino mass coming from the sneutrino mass 
splitting can also be significant \cite{grossman-haber1}. In our analysis, 
we have considered this loop effect so that the contribution to 
$(m_\nu)_{ee}$ comes from both tree and one-loop level. Writing this total
contribution as $(m_\nu)_{ee} = (m_\nu)^0_{ee} + (m_\nu)^1_{ee}$, we use the 
constraint $|(m_\nu)_{ee}| <$ 0.2 eV. Here, $(m_\nu)^0_{ee}$ is the 
tree-level value discussed in Eq. (\ref{eqn-deltamsnu}) and $(m_\nu)^1_{ee}$ 
is the one-loop contribution. We will also discuss how far below we can go with 
$|(m_\nu)_{ee}|$ so that the signal significance is $\ge 5 \sigma$. In 
order to show an example of the strength of this one-loop contribution, 
let us choose a sample point in the parameter space : $m_{3/2}$ = 50 TeV, 
$m_0$ = 250 GeV, $\tan\beta$ = 7 and $\mu <$ 0. If we now choose the tree-level value 
$({m_\nu})^0_{ee}$ = 0.079 eV, then the loop contribution is $(m_\nu)^1_{ee}$ 
$\approx$ 0.117 eV and the total contribution is consistent with the bound of 0.2 eV. 
In this case, the sneutrino mass splitting is $\approx$ 127 eV.

\begin{figure}
\vspace*{-3in}
\includegraphics{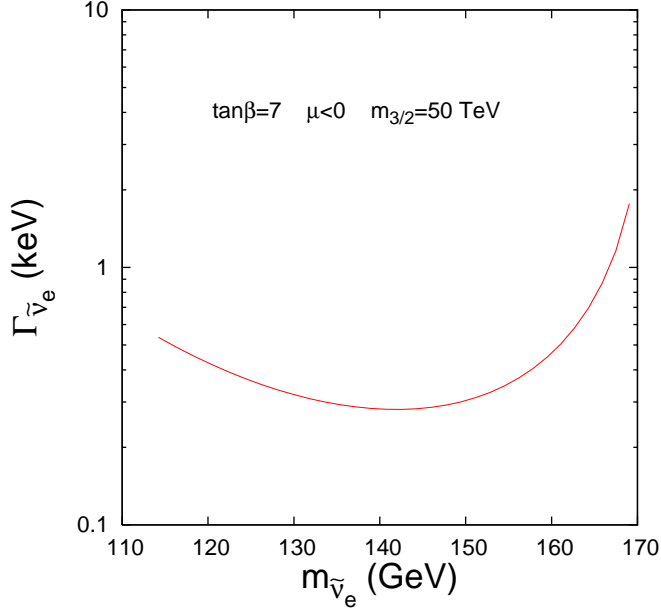}
\caption{\label{fig1}
Total decay width of the electron-sneutrino (${\tilde \nu}_e$) as a 
function of the sneutrino mass. Here, we have fixed $m_{3/2} = 50$ TeV and 
$\tan\beta= 7$. The sign of $\mu$ is taken to be negative.}
\end{figure}
\begin{figure}
\vspace*{-3in}
\includegraphics{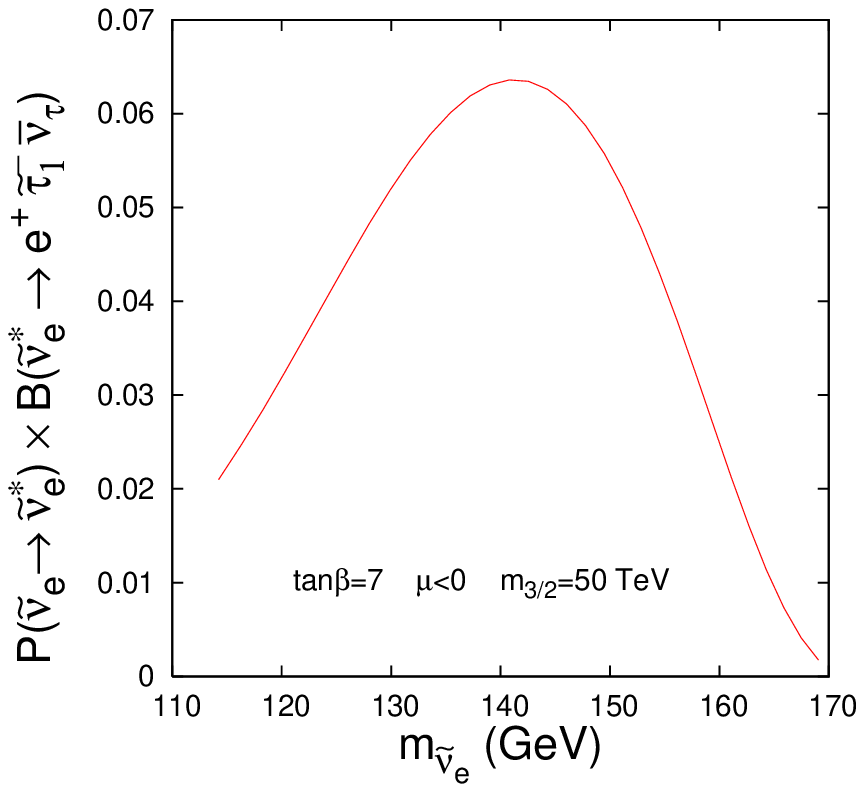}
\caption{\label{fig2}
The probability of observing a positron in the decay of the ${\tilde 
\nu}_e$ (defined in Eq. (\ref{eqn-prob})) as a function of $m_{{\tilde 
\nu}_e}$. Other parameter choices are the same as in Fig. \ref{fig1}.}
\end{figure}

Let us now show plots of the total decay width of the sneutrino/antisneutrino
($\Gamma_{\tilde \nu}$) and the probability of observing an opposite sign 
charged lepton in the final state of the decay of the 
sneutrino/antisneutrino given by Eq. (\ref{eqn-prob}). We have plotted 
these quantities for a fixed value of $m_{3/2}$ and $\tan\beta$, and the 
sign of $\mu$ is negative. The value of $m_0$ is changed in such a way 
that the condition (\ref{spectrum}) is satisfied. In Fig. \ref{fig1}, we 
have plotted the total decay width of the ${\tilde \nu}_e$ as a function 
of the sneutrino mass. The total width is calculated for the available 
three-body decay modes ${\tilde \nu}_e \rightarrow e^- {\tilde \tau}_1^+ 
\nu_\tau$ and ${\tilde \nu}_e \rightarrow \nu_e {\tilde \tau}_1^\pm \tau^\mp$ 
mediated by virtual charginos and neutralinos, respectively. It is worth 
mentioning here that the neutralino-mediated modes ${\tilde \nu}_e 
\rightarrow \nu_e {\tilde \tau}_1^- \tau^+$ and ${\tilde \nu}_e 
\rightarrow \nu_e {\tilde \tau}_1^+ \tau^-$ can, in general, have quite 
different partial decay widths. From Fig. \ref{fig1}, we can see that the 
total decay width is $\sim$ a few hundreds of eV in this region of the 
parameter space, being consistent with the requirement of observing a 
sizeable signal. In Fig. \ref{fig2}, we have plotted the probability of 
observing a positron in the decay of the ${\tilde \nu}_e$ (defined in 
Eq. \ref{eqn-prob}) as a function of the mass of the ${\tilde \nu}_e$ for 
the same choice of parameters as in Fig. \ref{fig1}. The 
probability shows a peak for $m_{{\tilde \nu}_e} \approx 140$ GeV, since  
the total decay width $\Gamma_{{\tilde \nu}_e}$ is smallest for this value 
of $m_{{\tilde \nu}_e}$, and, hence, the quantity $x_{\tilde \nu}$ defined 
in Eq. (\ref{xnu}) is largest for a fixed $\Delta m_{\tilde \nu}$. On the 
other hand, the branching ratio of ${\tilde 
\nu}_e^* \rightarrow e^+ {\tilde \tau}_1^- {\bar \nu_\tau}$ does not 
change much for the range of $m_{{\tilde \nu}_e}$ shown in the figure. We 
can also see that the probability is not so small for this choice of 
parameters, and if the production cross section $e^-\gamma \rightarrow 
{\tilde \chi}_1^- {\tilde \nu}_e$ is large, a sizeable number of positron 
events can be seen. 

Let us then move to discussing the physics of the electron-photon 
colliders and the production cross section of $\tilde \chi^-_1 {\tilde \nu}_e$ with
polarized and unpolarized beams. We will show the available region of 
the parameter space of the AMSB model where a reasonable number of signal 
events can be seen, while satisfying various experimental constraints. In 
addition, we will discuss the conditions under which the long-lived 
${\tilde \tau}_1$s produce heavily ionizing charged tracks before decaying 
eventually to lepton and neutrino pairs so that they can be easily 
distinguished from other sleptons.

\section{$e^- \gamma$ collider and the photon spectrum}

The way to obtain very high energy photon beams is to induce laser back-scattering off an
energetic $e^\pm$ beam \cite{egammacollider}. The reflected photon beam 
carries off only a fraction ($y$) of the energy of $e^\pm$ with
\bea
y_{max} = \frac {z} {1+z}, ~~~~z\equiv {\frac{4E_b E_L}{m^2_e}}{\rm cos}^2 {\frac 
{\theta_{bL}}{2}},
\eea   
where $E_{b(L)}$ are the energies of the incident electron/positron beam 
and the laser, respectively, and $\theta_{bL}$ is the incidence angle. The 
energy of the photon can be increased, in principle, by increasing the 
energy of the laser beam. However, a large $E_L$ (or, equivalently a large 
$z$) also enhances the probability of $e^+ e^-$ pair
creation through interactions between the laser and the scattered photon, 
consequently resulting in beam degradation. An optimal choice is 
$z=2(1+\sqrt{2})$, and this is the value we use in our analysis. The use of perfectly 
polarized electron and photon beams maximizes the signal cross section, though, 
in reality, it is almost impossible to achieve perfect polarizations. It is also 
extremely unlikely to have even near monochromatic high energy photon beams.  

For an $e^-\gamma$ collider, the cross sections can be obtained 
by convoluting the fixed-energy cross sections ${\hat \sigma}({\hat s}, 
P_\gamma, P_{e^-})$ with the appropriate photon spectrum:
\bea
\sigma(s) = \int {\mathrm dy} {\mathrm d{\hat s}}{\frac {\mathrm dn}{\mathrm dy}}
(P_b,P_L){\hat \sigma} ({\hat s}, P_\gamma, P_{e^-}) \delta({\hat s} - ys),
\eea   
where the photon polarization $P_\gamma$ is a function of $P_{b,L}$ and the momentum
fraction $y$ through the relation $P_\gamma = {P_\gamma}(y,P_b,P_L)$. In 
our analysis, we shall, for simplicity, consider circularly polarized 
laser beam scattering off polarized electron(positron) beams. The 
corresponding number-density $n(y)$ and average helicity
for the scattered photons are then given by \cite{egammacollider,egamma-tdr}
\bea
{\frac {dn}{dy}} &=& {\frac {2\pi\alpha^2}{m^2_e z \sigma_C}}C(y), \nonumber \\
{P_\gamma(y)}&=&{\frac{1}{C(y)}}\left\lbrack P_b \left\{{\frac {y}{1-y}}+y(2r -1)^2 
\right\} -P_L(2r-1) \left(1-y+{\frac {1}{1-y}}\right) \right\rbrack, \nonumber \\
C(y) &\equiv&Â{\frac {1}{1-y}} + (1-y) - 4r (1-r)-P_bP_Lrz(2r-1)(2-y),
\label{eqphotonspec}
\eea
where $r \equiv {y \over {z(1-y)}}$, and the total Compton cross section 
$\sigma_C$ provides the normalization.

It is also important to address another experimental issue regarding the long low-energy
tail of the photon spectrum. In a realistic situation \cite{egamma-tdr}, it is 
possible that these low-energy photons might not participate in any interaction. 
The harder back-scattered photons are emitted at smaller angles with respect to 
the direction of the initial electron, whereas softer photons are emitted at 
larger angles. Since the photons are distributed according to an effective spectrum 
(Eq. (\ref{eqphotonspec})), the low-energy photons which are produced at a wide 
angle are essentially thrown out, since they do not contribute significantly to any 
interaction. However, the exact profile of this effective spectrum is not simple, 
and it depends somewhat on the distance between the interaction point and the point 
where the laser photons are back-scattered and on the shape of the electron beam. 
Unfortunately, we are not in a position to include this effect in our simulations. 
It has been indicated in \cite{chou-cuypers} that neglecting this effect does not 
change the total signal cross section to any significant extent. 

Perfect polarization is relatively easy to obtain for the laser beam, and 
we shall use $|P_L| = 1$. However, the same is not true for electrons or 
positrons, and  we use $|P_b| = |P_{e^-}| = 0.8$ as a conservative choice. 
Since we want to produce the sneutrino in this study, the $e^-$ should be 
left-polarized, i.e. $P_{e^-} = -0.8$. In order to  improve the 
monochromaticity of the outgoing photons, the laser and the 
$e^\pm$ beam should be oppositely polarized \cite{berge}, which means $P_L 
\times P_b <$ 0. In our analysis, we shall use both choices of
polarizations consistent with $P_L \times P_b <$ 0.  

\section{Signal and backgrounds}

As explained in the Introduction, we will focus on the production process 
\cite{egamma-prod} $e^- \gamma \to {\tilde \nu}_e {\tilde \chi^-_1}$ and then look 
at the oscillation of the ${\tilde \nu}_e$ into a ${\tilde \nu}^*_e$. 
The resulting antisneutrino then decays through the three-body channel ${\tilde \nu}^*_e 
\to e^+ {\tilde \tau}_1^- {\bar \nu}_\tau$ with a large branching ratio. The chargino 
$\tilde \chi^-_1$ subsequently decays into a ${\tilde \tau}_1^-$ and an 
antineutrino ($\bar{\nu}_\tau$). The neutrinos escape detection and give rise 
to an imbalance in momentum. 
The signal is then
\bea
e^- \gamma \to {\tilde \nu}_e {\tilde \chi^-_1} \to e^+ + {\tilde \tau}_1^- + {\tilde
\tau}_1^- + \mpT,
\label{signal}
\eea   
where the two ${\tilde \tau}_1^-$s are long-lived and can produce heavily 
ionizing charged tracks inside the detector after traversing a macroscopic 
distance. The positron serves as the trigger for the event. The 
probability that the chargino decays before travelling a distance $\delta$ 
is given by $P(\delta) = 1-{\rm exp}(-\delta/L)$, where $L$ is the 
average decay length of the chargino. We assume that the ${\tilde \tau}_1^-$ decays 
through a tiny $R$-parity-violating coupling \cite{rpv-review} 
$\lambda_{233} = 5 \times 10^{-9}$ into charged lepton + neutrino pairs so that 
a substantial number of events do have a reasonably large decay lengths for 
which the displaced vertex  may be visible. At the end of this section, we will 
discuss the possible modifications in the signal event in order to accommodate 
a larger $R$-parity-violating coupling that will allow faster decay rates of 
the ${\tilde \tau}_1^-$s. Obviously, in such a situation, the Standard Model 
(SM) backgrounds would arise. We shall give numerical estimates of these SM 
backgrounds and discuss their implications. 

The cross section of the signal event in Eq.(\ref{signal}) has been 
calculated in the narrow width approximation. We have calculated the 
$2\to 2$ differential cross section ${d\sigma(e^- \gamma \to {\tilde
\nu}_e {\tilde \chi^-_1})}/ {d\cos\theta}$ and then folded into it the
probability of the sneutrino oscillation and proper branching fractions of the 
corresponding decay channels mentioned earlier to get the final state described 
above. 

We select the signal events in Eq. (\ref{signal}) according to the 
following criteria$\colon$ \\
$\bullet$ The transverse momentum of the positron must be large enough : 
$p^{e^+}_T> 10$ GeV. \\
$\bullet$ The transverse momentum of the ${\tilde \tau}_1^-$s must satisfy 
$p^{{\tilde \tau}_1}_T> 10$ GeV. \\
$\bullet$ The positron and both the staus must be relatively  central, i.e. their
pseudorapidities must fall in the range $|\eta^{{e^+},{\tilde \tau}_1}|< 2.5$. \\
$\bullet$ The positron and the staus must be well-separated from each other$\colon$
i.e. the isolation variable $\Delta R \equiv \sqrt{(\Delta \eta)^2 + (\Delta \phi)^2}$
(where $\eta$ and $\phi$ denote the separation in rapidity and the 
azimuthal angle, respectively) should satisfy $\Delta R> 0.4$ for each 
combination. \\
$\bullet$ The missing transverse momentum $\mpT > 10$ GeV. \\
$\bullet$ Both the heavily ionizing charged tracks due to the long-lived staus should
have a length $\ge 5$ cm. 

\begin{figure}
\includegraphics{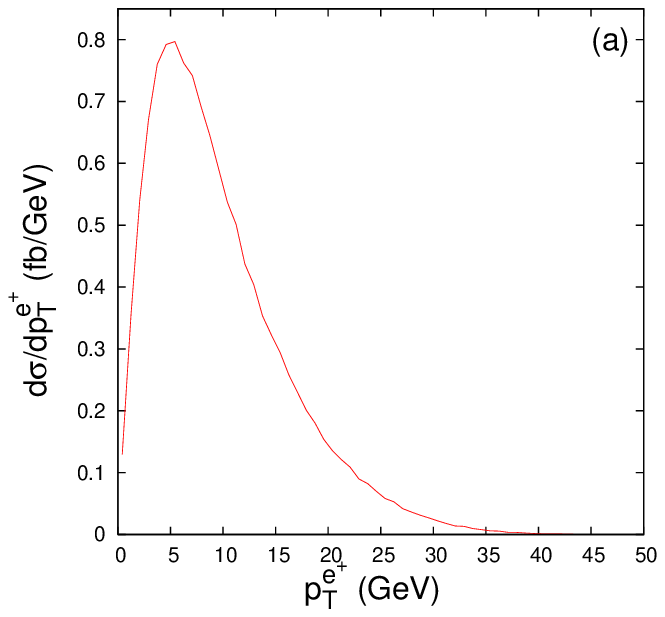}
\includegraphics{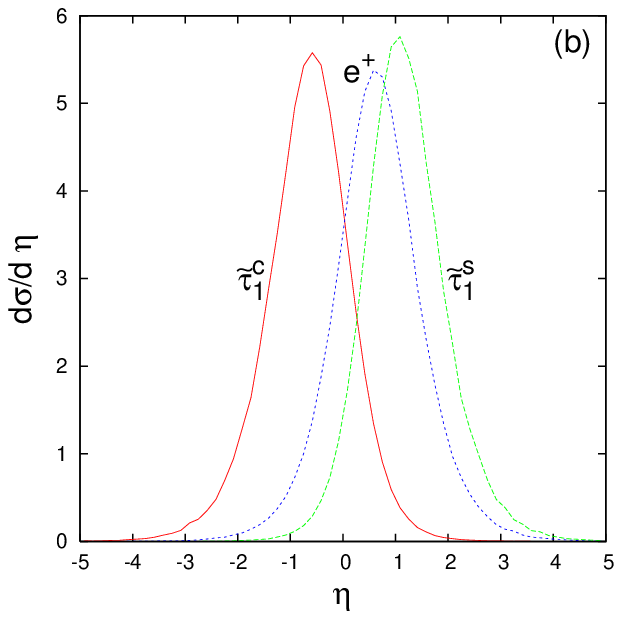}
\includegraphics{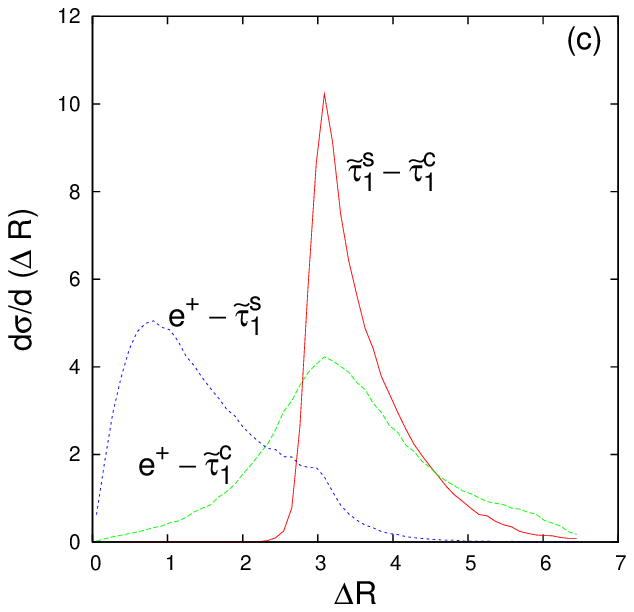}
\caption{\label{fig3} Kinematic distributions of signal events at a machine operating 
with $\sqrt{s_{ee}}$ = 500 GeV. The different plots correspond to (a) positron $p_T$; 
(b) rapidities; and (c) cone separations. Here, ${\tilde \tau}_1^s$ is the stau
arising from the decay of the antisneutrino and ${\tilde \tau}_1^c$ is the other stau
arising from the decay of the chargino. The choice of the beam polarization 
and other parameters are described in the text.}   
\end{figure}

\subsection{The signal profile}

In order to understand the profile of the signal we are looking for and to 
see the effects of the cuts we have employed, it is important to look 
at the kinematic distributions of various quantities. We will illustrate 
this for a sample point in the parameter space : $m_0 = 250$ GeV, 
$m_{3/2} = 50$ TeV, $\tan\beta = 7$ and $\mu < 0$, leading to ($m_{\tilde \chi^\pm_1}$, 
$m_{{\tilde \nu}_e}$, $m_{{\tilde \tau}_1}$) $\approx$ (170, 142, 122) GeV. Beam 
polarization choices are $P_L = +1$, $P_b = P_{e^-} = -0.8$. As explained earlier,
these distributions have been calculated in the narrow width approximation by 
convoluting the distributions of the ``wrong-sign charged lepton" with those of the
sneutrino from the $2\to2$ process $e^- \gamma \to {\tilde \nu}_e {\tilde 
\chi^-_1}$. For such a choice of parameters, the total cross section is 
$\approx 10.26$ fb without any cuts for a machine operating at $\sqrt{s_{ee}}=500$ GeV. The transverse 
momentum of the positron in Fig. \ref{fig3}(a) shows its peak around 5 GeV and then 
falls sharply. This is due to the fact that the positron is coming from the decay 
of the ${\tilde \nu}_e^*$. Since the mass difference between the ${\tilde \nu}_e^*$ 
and the ${\tilde \tau}_1$ is not large, most of the positrons are softer. Hence, the 
requirement of the minimum positron transverse momentum of 10 GeV rejects quite a few 
signal events. However, this $p_T$ cut of 10 GeV is needed in order to trigger the 
event. Since the positron is coming out of the decay of the
antisneutrino, it is quite central, which can also be seen from its 
rapidity distribution in Fig. \ref{fig3}(b). On the other hand, one of the 
staus comes from the decay of the chargino and the other one comes from 
the decay of the sneutrino. The one (${\tilde \tau}_1^s$) arising from the sneutrino 
decay shows quite similar behaviour as the positron, whereas the other (${\tilde
\tau}_1^c$) arising from the chargino decay, though being quite central, is somewhat
boosted in the opposite direction (Fig. \ref{fig3}(b)). One can immediately understand
that the ${\tilde \tau}_1^c$ is well-separated from the positron and the 
other stau (${\tilde \tau}_1^s$). This feature is evident from the 
distribution shown in Fig. \ref{fig3}(c). This conclusion holds almost over the 
entire allowed parameter space which we are considering. On the other hand, 
the angular separation between the positron and ${\tilde \tau}_1^s$ is much smaller, 
and the position of the peak moves slightly depending on the point chosen in the 
allowed parameter space. Because of the choice of a very small $R$-parity-violating 
coupling, both the staus leave a substantial track, and, for most of the events, 
the decay lengths are greater than 10 cm. Some of the events can have charged 
tracks which may extend up to a meter or greater than that.

\subsection{The SUSY backgrounds}

As one can see, due to the presence of these heavily ionizing charged tracks, the 
signal is entirely free of any Standard Model (SM) background. 
However, there are backgrounds from SUSY processes. 
One possibility is the associated production 
of the left-selectron (${\tilde e}_L^-$) and the lightest neutralino (${\tilde 
\chi}^0_1$). If the mass of the left-selectron is larger than the mass of the lighter 
chargino, then it may subsequently decay into an ($e^-$ + ${\tilde 
\chi}^0_1$) pair or a (${\tilde \chi}^-_1$ + $\nu_e$) pair with the 
branching fractions $B({\tilde e}_L^- \rightarrow e^- + {\tilde \chi}^0_1) 
\approx 33-39$ \% and $B({\tilde e}_L^- \rightarrow \nu_e + {\tilde 
\chi}^-_1) \approx 60-66$ \%, due to the fact that both ${\tilde 
\chi}^0_1$ and ${\tilde \chi}^-_1$ are predominantly winos. Two-body decays into 
heavier neutralinos (charginos) are kinematically forbidden 
for most of the parameter space. In order to have the same visible final 
state as in the case of our signal, we concentrate on the 
neutrino-chargino channel. The lightest neutralino (${\tilde \chi}^0_1$) 
may decay into an electron-neutrino ($\nu_e$) and the associated 
antisneutrino (${\tilde \nu}_e^*$). Due to the choice of our spectrum in 
Eq. (\ref{spectrum}), ${\tilde \chi}^0_1$ always has this two-body mode 
available. The resulting antisneutrino can go to the three-body channel 
${\tilde \nu}_e^* \rightarrow e^+ {\tilde \tau}^-_1 {{\bar \nu}_\tau}$, and the 
chargino arising from the selectron decay may go into a (${\tilde \tau}_1^- + {\bar 
\nu}_\tau$) pair. The final background event is then $e^- \gamma 
\rightarrow {\tilde e}_L^- {\tilde \chi^0_1} \rightarrow e^+ {\tilde 
\tau}^-_1 {\tilde \tau}^-_1 {\nu_e} {{\bar \nu}_\tau} {\nu_e} {{\bar \nu}_\tau}$, 
where the neutrinos give rise to the missing transverse momentum $\mpT$.  

In order to compare the strength of the background and the signal event, 
let us give an example. For $m_0 = 255$ GeV, $m_{3/2} = 50$ TeV, 
$\tan\beta = 7$ and $\mu < 0$, the spectrum is $m_{{\tilde e}_L} = 171.4$ 
GeV, $m_{{\tilde \nu}_e} = 151.1$ GeV, $m_{{\tilde \chi}_1^-} = 170.1$ GeV, 
$m_{{\tilde \chi}_1^0} = 169.94$ GeV and $m_{{\tilde \tau}_1} = 131.8$ GeV. After 
imposing our cuts at $\sqrt{s_{ee}}=500$ GeV, the surviving background is 
0.48 fb and the signal is 2.33 fb. Here, the polarization choices are the same as in
the previous subsection. If we now calculate the signal significance 
$\equiv$ $N_e/\sqrt{N_e+N_B}$, where $N_e$ is the number of signal events 
and $N_B$ is the number of background events, then, for this particular example, 
the ratio is much greater than 5 for an integrated luminosity of 
500 ${\rm fb}^{-1}$. If we increase the value of $m_0$, 
then the masses of both the ${\tilde e}_L$ and the ${\tilde \nu}_e$ increase, and, as 
a result, the signal as well as the background cross sections decrease. However, the 
signal significance always remains greater than 5. On the other hand, if we keep 
$m_0$ fixed and change the value of $m_{3/2}$ in such a way that ${\tilde e}_L$ is 
always heavier than ${\tilde \chi}_1^-$ and ${\tilde \chi}_1^0$, then the signal 
significance remains again greater than 5. One of the reasons for this small cross 
section of the background event is that the branching ratio $B({\tilde \chi}_1^0 
\rightarrow \nu_e {\tilde \nu}_e^*$) is very small (less than 10 \%). It is worth 
mentioning that the decay ${\tilde \chi}_1^0 \rightarrow {\bar \nu}_e {\tilde \nu}_e$ 
could contribute to the signal through the ${\tilde \nu}_e$--${\tilde \nu}_e^*$ 
oscillation, but this process is further suppressed by the small oscillation 
probability (less than 0.1) and is hence negligible. 

In the case when ${\tilde e}_L^-$ is lighter than ${\tilde \chi}_1^-$ and 
${\tilde \chi}_1^0$, it can decay into the chargino-mediated three-body 
mode ${\tilde e}_L^- \rightarrow \nu_e {\tilde \tau}_1^- {\bar 
\nu}_{\tau}$, which contributes to the background. In this case, one 
should notice that ${\tilde \chi}_1^0$ goes to the two-body mode ${\tilde \nu}_e^* \nu_e$ 
with a branching ratio smaller than in the earlier case. There are other 
three-body decay modes available for the left-selectron in this case, namely, 
${\tilde e}_L^- \rightarrow e^-\tau^\pm {\tilde \tau}_1^\mp$, ${\tilde e}_L^- 
\rightarrow  {\nu_e} {\ell^-}{\tilde\nu}^*_\ell$, ${\tilde e}_L^- \rightarrow e^- 
{\nu_{\mu,\tau}} {{\tilde \nu}_{\mu,\tau}^*}, e^- {\bar \nu_{\mu,\tau}} 
{{\tilde \nu}_{\mu,\tau}}$ and ${\tilde e}_L^- \rightarrow {\tilde \nu}_e \ell^- 
{\bar \nu}_\ell$ where $\ell = e, \mu, \tau$. Here, we have neglected the three-body 
decays involving ${\tilde e}_R$ and ${\tilde \mu}_R$ in the final state. From the above 
discussion, we can conclude that the cross section for the background 
event still remains quite small in the case when ${\tilde e}_L$ is lighter 
than ${\tilde \chi}_1^-$ and ${\tilde \chi}_1^0$. On the other hand, the signal also 
suffers a suppression due to a smaller branching ratio of ${\tilde \chi}^-_1$ in the 
(${\tilde \tau}_1^- + \nu_\tau$) mode. However, this suppression is such that the 
ratio $N_e/\sqrt{N_e+N_B}$ always remains greater than or equal to 5.     

Another source of background could be the associated production $e^- \gamma
\rightarrow {\tilde e}_L^- {\tilde \chi}^0_2$. However, this production 
process is kinematically forbidden for the entire region of the parameter 
space we are investigating for a machine operating at 
$\sqrt{s_{ee}}$ = 500 GeV. For a $\sqrt{s_{ee}}$ = 1 TeV collider, 
this process is allowed, but the production cross section is too small to 
contribute significantly. The 2 $\rightarrow$ 3 process
$e^- \gamma \rightarrow {\nu_e} {{\tilde e}_L^-} {\tilde \nu}_e^*$ could also
contribute to the background, but the production cross section in this 
case is very small ($< {\mathcal O}(10^{-2})$ fb) \cite{like-sign-rp}.

\subsection{The signal strength and the parameter space}

Let us now discuss the signal event in more detail. The number of signal events
and the kinematical distributions depend crucially on the sneutrino and the
chargino masses and also on the mass of ${\tilde \tau}_1$. In our analysis, the
evolution of gauge and Yukawa couplings as well as that of scalar masses are
computed using two-loop renormalization group equations (RGE) \cite{martin-vaughn}.
We have also incorporated the unification of gauge couplings at the scale $M_G \sim
2 \times 10^{16}$ GeV with $\alpha_3(M_Z) \approx 0.118$. The boundary 
conditions for the scalar masses are given at the unification scale via 
Eq. (\ref{eqn-scalar}). The magnitude of the higgsino mass parameter $\mu$ 
is computed from the requirement of a radiative electroweak symmetry 
breaking and at the complete one-loop level of the effective potential 
\cite{potential}. The optimal choice of the renormalization scale is expressed 
in terms of the masses of the top-squarks through the relation $Q^2 =
m_{{\tilde t}_1} m_{{\tilde t}_2}$. We have also included the supersymmetric QCD
corrections to the bottom-quark mass \cite{susy-qcd}, which is significant for
large $\tan\beta$. It should be noted at this point that gaugino masses and
trilinear scalar couplings can be computed from the expressions in
Eqs. (\ref{eqn-gaugino}) and (\ref{eqn-trilinear}) at any scale once 
the appropriate values of the gauge and Yukawa couplings at that scale are 
known. A particularly interesting feature of the mAMSB model is that the 
lighter chargino ${\tilde \chi}^\pm_1$ and the lightest neutralino 
${\tilde \chi}^0_1$ are both almost exclusively a wino and, hence, nearly 
mass-degenerate. A small mass difference is generated from the tree-level 
gaugino-higgsino mixing as well as from the one-loop corrections to the 
chargino and the neutralino mass matrices \cite{mass-diff}. The numerical 
results of the spectrum of mAMSB model have been obtained using the 
fortran codes developed in \cite{muong-2-1} and in the first two 
references of \cite{amsb-linearee}. We have checked that our results agree with
those of previous authors \cite{amsb-hadron1} for a few sample choices of
parameters.

It is important to look at the total number of signal events as a function of the model
parameters with the condition on the spectrum given in Eq. 
(\ref{spectrum}). In order to do this, we will fix the value of 
$\tan\beta$ and take the signature of $\mu$ to be negative and then 
allow $m_0$ and $m_{3/2}$ to vary in a region which satisfies the experimental 
constraints on the sparticle masses. Later on, we will also discuss how 
the total cross section changes with $\tan\beta$ and the sign of $\mu$. As 
above, we make a specific choice for the beam polarization, namely, $P_L = +1$, 
$P_b = P_{e^-} = -0.8$. 

\begin{figure}
\vspace*{-3in}
\includegraphics{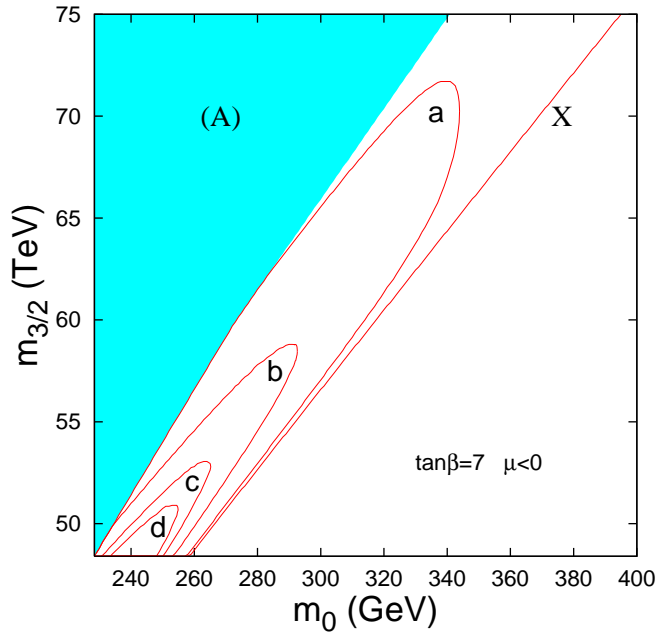}
\caption{\label{fig4} Parameter regions with $\tan\beta=7$ and $\mu<0$. The area (A)
represents the parameter region forbidden by the stau mass bound. The mass spectrum
(\ref{spectrum}) is obtained in the region between the area (A) and the line X.
Assuming an integrated luminosity of $500$ fb$^{-1}$ at $\sqrt{s_{ee}}=500$ GeV, 
the numbers of positron events per year inside the contours are: (a) $N_e \geq 50$, 
(b) $N_e \geq 500$, (c) $N_e \geq 1000$ and (d) $N_e \geq 1300$ for $(m_\nu)^0_{ee}$ 
= 0.079 eV so that the total contribution $(m_\nu)_{ee} \approx$ 0.2 eV, while 
satisfying $N_e \geq 5 \sqrt{N_e+N_\mathrm{B}}$.}
\end{figure}

In Fig. \ref{fig4}, we show our results for the total number of positron 
events for a machine operating at $\sqrt{s_{ee}}$ = 500 GeV with 500 fb$^{-1}$ integrated
luminosity, after imposing the kinematical cuts discussed above. The region marked by (A)
corresponds to a lighter stau mass of less than 86 GeV \cite{PDG}. The area below the 
line X does not satisfy the mass hierarchy of Eq. (\ref{spectrum}). Thus, 
the allowed region in the ($m_0-m_{3/2}$) plane is the one between the area (A) and 
the line X. The other experimental constraints \cite{PDG} which we have used are the 
mass of the lighter chargino ($m_{\tilde \chi^\pm_1} >$ 104 GeV), the mass of the 
sneutrino ($m_{\tilde \nu} >$ 94 GeV) and the mass of the lightest Higgs boson 
\cite{haber-carena} ($m_h >$ 113 GeV). Apart from these direct bounds, one 
should also consider the constraints on the parameter space arising from 
the virtual exchange contributions to low-energy observables. For example, 
the constraints on the minimal AMSB model parameters from the measurement of muon 
anomalous magnetic moment have been studied in several works \cite{muong-2-1, 
muong-2-2, muong-2-3}. However, the numerical results of those papers should be 
modified due to the reevaluation of the light by light hadronic contribution 
\cite{lightbylight} and the results published by the E821 experiment \cite{E821}. 
In addition, one should bear in mind that the theoretical calculation of the SM 
contribution to muon $(g-2)$ has many remaining theoretical uncertainties. The 
measurement of the rare decay $\Gamma(B \rightarrow X_s \gamma)$ can set additional 
bounds \cite{muong-2-2, muong-2-3} on the parameters, but they are not very 
restrictive. Bounds can also be obtained by demanding that 
the electroweak vacuum corresponds to the global minimum of the scalar potential 
\cite{vacuumstability}. However, as long as it can be ensured that the local minimum 
has a life time longer than the present age of the Universe, these additional bounds 
can be evaded \cite{universe}.    

We have used the value of $\tan\beta = 7$, and the sign of $\mu$ is taken 
to be negative. It has already been mentioned that, in the 
AMSB scenario, the positron events in an $e^-\gamma$ collider via the 
sneutrino-antisneutrino mixing can provide information on the neutrino 
mass matrix elements $(m_\nu)_{ee}$. In Fig. \ref{fig4}, we have chosen 
the value of $(m_\nu)_{ee}^0 = 0.079$ eV which corresponds to 
$(m_\nu)_{ee} \approx$ 0.2 eV. Later on, we will make comments on the 
smallest value of $(m_\nu)_{ee}$ which can be probed in this scenario. In 
this figure, we have plotted contours of total number of positron events 
$N_e$, starting with $N_e$ = 50. It is evident from this figure that an 
experiment of this type can easily explore $m_{3/2}$ as high as $\approx$ 72 TeV 
whereas the reach in $m_0$ is $\approx$ 340 GeV for a negative $\mu$, 
$\tan\beta$=7 and $(m_\nu)_{ee} \approx $ 0.2 eV. Even with an integrated luminosity of 100 fb$^{-1}$, 
it is possible to explore values of $m_{3/2}$ and $m_0$ up to $\approx$ 
64 TeV and 315 GeV, respectively, with $N_e \geq 50$.  
\begin{figure}
\vspace*{-3in}
\includegraphics{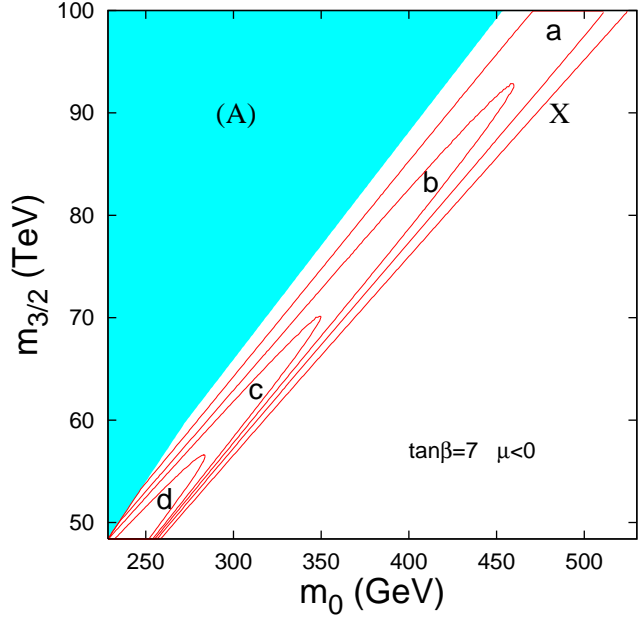}
\caption{\label{fig5} Parameter regions with $\tan\beta=7$ and $\mu<0$.
The area (A) represents the parameter region forbidden by the stau mass
bound. The mass spectrum (\ref{spectrum}) is obtained in the region
between the area (A) and the line X. Assuming an integrated luminosity of
$500$ fb$^{-1}$ at $\sqrt{s_{ee}}=1$ TeV, the numbers of positron events
per year inside the contours are: (a) $N_e \geq 100$, (b) $N_e \geq 200$,
(c) $N_e \geq 300$ and (d) $N_e \geq 500$ for $(m_\nu)^0_{ee}$ = 0.079 eV
so that the total contribution $(m_\nu)_{ee} \approx$ $0.2$ eV, while
satisfying $N_e \geq 5 \sqrt{N_e+N_\mathrm{B}}$.}
\end{figure}
In Fig. \ref{fig5}, we show a similar plot in the ($m_0$ -- $m_{3/2}$) plane for 
a machine operating at $\sqrt s$ = 1 TeV with other inputs remaining the same. We can 
see similar features in both the Figures \ref{fig4} and \ref{fig5}, with the obvious 
enhancement in the reach in the latter case.   

So far, we have discussed the strength of the signal for a fixed value of the 
parameter $\tan\beta$. Let us now see how the signal cross section changes 
with varying $\tan\beta$. For a fixed $m_{3/2}$ and with increasing 
$\tan\beta$, a larger value of $m_0$ is required to obtain the spectrum (\ref{spectrum}) 
for both choices of the sign of $\mu$. This leads to a larger value of 
$m_{{\tilde \nu}_e}$. On the other hand, $m_{{\tilde \tau}_1}$ becomes smaller due to 
a stronger left-right mixing in the stau sector. Thus, the ratio ${m_{{\tilde 
\nu}_e}}/{m_{{\tilde \tau}_1}}$ increases with $\tan\beta$, making the 
sneutrino decay width $\Gamma_{{\tilde \nu}_e}$ an increasing function of 
$\tan\beta$. Hence, in order to get a sizeable number of positron events 
$N_e$, one needs a small value of the parameter $\tan\beta$ for a fixed 
neutrino mass. This feature has also been observed in Ref. 
\cite{like-sign-norp}. For $\mu >$ 0, the Higgs boson mass is a bit lower, 
so we need a higher $\tan\beta$ for a fixed $m_{3/2}$ to satisfy the constraint on 
the light Higgs boson mass $m_h > 113$ GeV. This makes the number of positron events 
very small. When $\tan\beta$ is fixed, the allowed region in the $m_0$ -- $m_{3/2}$ 
plane is shortened for $\mu>0$, since a higher value of 
$m_{3/2}$ is required in order to satisfy the Higgs boson mass bound. 
For a negative $\mu$, the requirement of observing the signal with at 
least $5\sigma$ significance 
implies that the highest allowed value of $\tan\beta$ is $\approx 8.1$ 
with a machine operating at $\sqrt{s_{ee}} = 500$ GeV, and $\tan\beta$
is $\approx 8.0$ for a machine 
with $\sqrt{s_{ee}}$ = 1 TeV. The lowest allowed 
value of $\tan\beta$ for a fixed $m_{3/2}$ is limited by the Higgs boson 
mass bound. For a negative $\mu$, we can have $\tan\beta$ as low as 
$\approx 4.9$, which will still produce an acceptable Higgs boson mass 
and at least $5\sigma$ signal significance for $\sqrt{s_{ee}} = 500$ GeV. 
For a value of $\tan\beta$ as low as 4.9, it is notable that the value of 
$m_0$ and $m_{3/2}$ should be quite high ($m_0 \approx 375$ GeV and 
$m_{3/2} \approx 88$ TeV) in order to have enough signal events. For 
$\sqrt{s_{ee}} = 1$ TeV, the lowest allowed value of $\tan\beta$ is 
approximately the same. In order to have $5\sigma$ signal significance for 
a positive $\mu$, we must have $\sqrt{s_{ee}} = 1$ TeV in which case the 
$\tan\beta$ range is quite small $\approx$ $6.1$--$6.3$, while the value 
of $m_{3/2}$ is of the order of $100$ TeV and $m_0$ is in the range 
$460$--$590$ GeV.

\begin{table}
\begin{center}
\footnotesize
\begin{tabular}{|l||c|c|c||c|c|c||}
\hline
(~$m_0~{\rm (GeV)},~m_{3/2}~{\rm (TeV)},~\tan\beta$~) &
\multicolumn{3}{|c||}{\hspace*{1em} (250,~50,~7)}   &
\multicolumn{3}{|c||}{\hspace*{1em} (350,~70,~7)} \\
\hline
(~$m_{\tilde\chi^-_1}, m_{{\tilde\nu}_e}, m_{{\tilde\tau}_1}~)~{\rm (GeV)}$
&\multicolumn{3}{|c||}{(~170.1,~142.5,~121.9~)}
  &  \multicolumn{3}{|c||}{(~239.6,~208.9,~179.3)}\\
\hline
& \multicolumn{3}{|c||}{$(P_L, P_b, P_{e^-})$}
& \multicolumn{3}{|c||}{$(P_L, P_b, P_{e^-})$} \\
\cline{2-7}
& ($-, +, -$) & ($+, -, -$) & ($0, 0, 0$)
& ($-, +, -$) & ($+, -, -$) & ($0, 0, 0$) \\
\hline
Total $\sigma$ (without cuts) (fb) &  7.15 & 5.93  & 3.21  &  1.29 & 1.45 & 0.66 \\
\hline
Total $\sigma$ (with cuts) (fb) &  2.12  &  1.58 & 0.92 &  0.56 & 0.59 & 0.28\\
\hline
\end{tabular}
\end{center}
\caption{Illustrating the effects of various polarization choices on the signal
cross sections for two specimen points in the parameter space and
for $\sqrt{s_{ee}}= 1$ TeV. In either case, $\mu < 0$. Whenever nonzero, 
$|P_L| = 1$, $|P_b| = |P_{e^-}| = 0.8$.}
\label{table}
\end{table}

In order to discuss the effect of the beam polarizations, we choose two 
sample points in the parameter space and show the results in Table 
\ref{table} for a machine with $\sqrt{s_{ee}}$ = 1 TeV. One can see that the cross 
sections for polarized beams are larger than the unpolarized ones. The effect of the 
cuts can also be seen. Depending on the choice of the parameters, the kinematical cuts 
can reduce the number of events by more than 50 \%.

\begin{figure}
\includegraphics{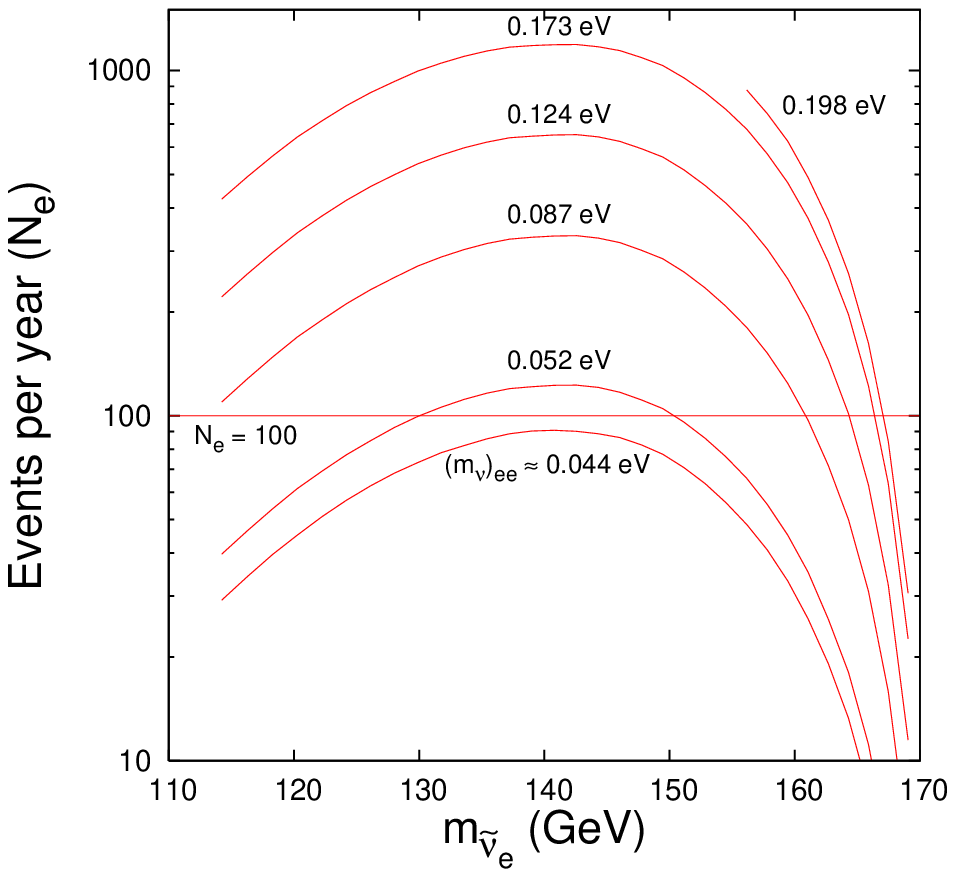}
\caption{\label{fig-mnu} Number of events ($N_e$) per year (with an integrated
luminosity of 500 $\mathrm{fb}^{-1}$) as a function of $m_{{\tilde
\nu}_e}$ for different choices of $(m_\nu)^0_{ee}$ as dicussed in the text. Here, 
$\tan\beta=7$, $\mu<0$ and $m_{3/2}$ = 50 TeV. The values of the total 
contribution $(m_\nu)_{ee}$ corresponding to each line are shown in the 
figure. The horizontal line stands for $N_e$ = 100 satisfying $N_e \geq 5
\sqrt{N_e+N_\mathrm{B}}$.}
\end{figure}

Let us now discuss the change in the number of events when $(m_\nu)^0_{ee}$ is
varied in such a way that it is consistent with the upper limit of $0.2$ eV
for the total contribution $(m_\nu)_{ee}$. For the purpose of this discussion, 
we choose a machine operating at $\sqrt{s_{ee}}$ = 500 GeV. It is evident 
from our discussion so far that larger values of $(m_\nu)^0_{ee}$ give 
a larger cross section. This is also shown in Fig. \ref{fig-mnu} for a 
sample choice of $m_{3/2}$ = 50 TeV, $\tan\beta$ = 7 and $\mu <$ 0. 
Assuming an integrated luminosity of 500 $\mathrm{fb}^{-1}$, we have
plotted the number of events per year as a function of $m_{{\tilde \nu}_e}$ for 
different choices of $(m_\nu)^0_{ee}$. The curves from below correspond to
$(m_\nu)^0_{ee}$ = 0.018 eV, 0.021 eV, 0.035 eV, 0.05 eV, 0.07 eV and 0.081 eV. The 
corresponding values of the total contribution $(m_\nu)_{ee}$ are shown in the figure. 
The horizontal line gives $N_e$ = 100 per year. This figure tells us that if we 
demand the value of $N_e$ to be $\geq$ 100, so that the signal significance is
$\geq$ 5$\sigma$, then we can probe the value of $(m_\nu)_{ee}$ down to $\approx$ 
0.05 eV. On the other hand, the current upper limit of 0.2 eV on $(m_\nu)_{ee}$ sets 
the upper limit of $(m_\nu)^0_{ee} \approx$ 0.081 eV. The topmost curve in this 
figure starts from a slightly higher value of $m_{{\tilde \nu}_e}$, since the bound 
on $(m_\nu)_{ee}$ is not satisfied before that. This figure can also be used to 
extract the value of $(m_\nu)_{ee}$ with the knowledge of the number of events 
and other masses.  

Finally, we will discuss the situation when a larger $R$-parity-violating 
(RPV) coupling is present. We shall assume that a single RPV coupling is 
dominant at a time, and our choice is $\lambda_{233}$. The reason behind 
this choice is that it will not affect the total decay width 
of ${\tilde \nu}_e^*$, since otherwise the $\Delta L$ = 2 effect will be 
diluted. The coupling $\lambda_{233}$ also allows observation of muons and taus 
in the final state from the decay of the ${\tilde \tau}_1^-$s, so that they can 
be clearly distinguished (assuming 100$\%$ detection efficiency) from the isolated 
$e^+$ produced due to the sneutrino mixing phenomena. The upper bound on the 
coupling $\lambda_{233}$ is given by \cite{rpv-bounds}
\bea
|\lambda_{233}| < 0.070 \times \frac {m_{{\tilde \tau}_R}}{100~{\mathrm {GeV}}}.
\label{lambda-bound}
\eea
The bound in Eq. (\ref{lambda-bound}) has been obtained from
measurements of $R_\tau = Â\Gamma(\tau \rightarrow e \nu \bar \nu)/\Gamma(\tau
\rightarrow \mu \nu \bar \nu)$ and $R_{\tau \mu} = \Gamma(\tau \rightarrow \mu \nu
\bar \nu)/\Gamma(\mu \rightarrow e \nu \bar \nu)$. Using this upper limit on
$\lambda_{233}$, we see that the ${\tilde \tau}_1^-$s will decay promptly to
either a $(\tau + \nu_\mu)$ pair or a $(\mu + \nu_\tau)$ pair. Taking into account
all possible final states involving $\mu$ and/or $\tau$, the signal event in this
situation looks like Â$e^- \gamma \rightarrow e^+ \ell^- \ell^- + \mpT$ where $\ell =
\mu, \tau$. 

The SM background to this process arises from the resonant production of three 
$W^\pm$ bosons through the $2 \rightarrow 4$ process $e^- \gamma \rightarrow W^-
W^- W^+ \nu_e$ and the subsequent decays of the $W^\pm$s. The background to the
signal $e^- \gamma \rightarrow e^+ \ell^- \ell^- + \mpT$ originates when the $W^+$
decays through $W^+ \rightarrow e^+ \nu_e$ and the two $W^-$s decay 
through $W^- \rightarrow \ell^- {\bar \nu}_\ell$. This process has already 
been calculated in Ref. \cite{pilaftsis} for two values of the c.m. 
energy, namely, $\sqrt{s_{\mathrm {ee}}}$ = 500 GeV and $\sqrt{s_{\mathrm {ee}}}$ = 1 TeV. As can
be seen from the Table 1 of Ref. \cite{pilaftsis}, the background cross 
section for $\sqrt{s_{\mathrm {ee}}}$ = 500 GeV is $\approx$ 0.0089 fb 
(after dividing the number in the table by $B(W^-W^- \rightarrow hadrons)$ 
and multiplying by $B(W^-W^- \rightarrow (\mu + \tau))$, 
and that for $\sqrt{s_{\mathrm {ee}}} = 1$ TeV is $\approx 0.1257$ fb. One 
should note here that these numbers for the background cross sections are 
without any cuts and should be reduced further after imposing suitable 
kinematical cuts. It should be mentioned here that the SUSY
background analysis remains almost the same as discussed in the subsection 
IVB but now with prompt decays of the ${\tilde \tau}_1^-$s.

In order to look at the signal to background ratio in this case, let us choose 
$\sqrt{s_{ee}} = 500$ GeV and two widely separated points in the parameter 
space$\colon$ 

\noindent (A) $m_0 = 255$ GeV, $m_{3/2} = 50$ TeV, $\tan\beta = 7$ and 
$\mu < 0$; and \\
(B) $m_0 = 310$ GeV, $m_{3/2} = 60$ TeV, $\tan\beta = 7$ and $\mu < 0$. 

The polarization choices are the same as in the subsection IVA.  
The signal cross section for the point (A) is 3.3 fb (with only a cut on 
the positron $p_T$, which is the most effective one, and on the positron 
pseudorapidity) and the SUSY background cross section is 0.67 fb. Combining 
this background cross section with the SM background mentioned above, we see 
that the signal to background ratio is greater than $5\sigma$. Similarly, for 
the point (B), the signal cross section is 0.425 fb and the SUSY background 
cross section is 0.328 fb. Again, we see that the ratio $N_e/\sqrt{N_e + N_B}$ 
is greater than 5. Thus, even in the case of a larger RPV coupling, we can 
explore an appreciable region in the parameter space of our interest with the 
signal described above. We have also performed a similar analysis for a 
$\sqrt{s_{ee}}$ = 1 TeV collider, and, once again, it shows that a significant 
region in the parameter space of mAMSB model with a spectrum given in Eq. 
(\ref{spectrum}) can be probed by this signal.

\section{Conclusion}

In conclusion, we have discussed the potential of an electron-photon 
collider to investigate the signature of $\tilde\nu_e$--${\tilde\nu}_e^*$ 
mixing in an AMSB model which can accommodate $\Delta L = 2$ Majorana 
neutrino masses. A very interesting feature of such models is that the 
sneutrino-antisneutrino mass splitting $\Delta m_{\tilde\nu}$ is
naturally large and is ${\mathcal O} ({4\pi{m_\nu}}/\alpha)$. On the other hand, 
the total decay width of the sneutrino is sufficiently small in a significant 
region of the allowed parameter space of the model. These two features enhance 
the possibility of observing sneutrino oscillation signal in various colliders. 
We have demonstrated that the associated production of the lighter chargino and 
the sneutrino at an $e^- \gamma$ collider could provide a very clean signature 
of such a scenario. 

The signal event consists of an energetic positron (resulting from the 
oscillation of a ${\tilde\nu}_e$ into a ${\tilde\nu}_e^*$), which serves 
as the trigger, two macroscopic heavily ionizing charged tracks in the 
detector coming from the long-lived staus and a large missing transverse momentum. 
Due to the presence of these macroscopic charged tracks
in combination with the energetic positron, the signal is free of any 
Standard Model backgrounds. The backgrounds from supersymmetric processes 
are present, but they are small and become even smaller with the cuts we have 
imposed. Consequently, with an integrated luminosity of 500 fb$^{-1}$, one could see 
as many as 1300 signal events in some region of the parameter space for a machine 
operating at $\sqrt{s_{ee}}$ = 500 GeV with polarized beams. We have also seen that 
the signal significance is $\ge 5\sigma$ for almost the entire allowed region of 
parameter space. In the case of a $\sqrt{s_{ee}}$ = 1 TeV collider, the features are
similar with an obvious enhancement in the reach. The signal cross section depends also 
on $(m_\nu)_{ee}$, and, obviously, we get the best result with the value 
of $(m_\nu)_{ee}$ (including both tree and one-loop contribution) close to its present 
upper limit. We have also discussed the effects on the signal cross section when 
lowering the value of $(m_\nu)_{ee}$. This way, the signal discussed in this paper 
can be used to determine $(m_\nu)_{ee}$ which provides important information on a 
particular combination of the neutrino masses and mixing angles, which is not possible 
to obtain from neutrino oscillation experiments. Slightly lower values of $\tan\beta$ 
($\lsim 9$) and a negative $\mu$ are preferred to get a sizeable number 
of signal events. Taking into account various experimental constraints and demanding 
that the signal significance should be $\geq 5\sigma$, we see that the lower limit on 
$\tan\beta$ is $\approx 4.9$. We have also discussed the possible effects on the
signal when a larger $R$-parity-violating coupling is introduced. 
Numerical estimates of the Standard Model backgrounds in this case have 
also been provided.    

\begin{acknowledgments}
We thank Dilip Kumar Ghosh for helpful discussions. This work is supported 
by the Academy of Finland (Project numbers 104368 and 54023). 
\end{acknowledgments}


\begin{thebibliography}{99}
\bibitem{altarelli-kayser}
For recent reviews on neutrino physics, see, e.g. G. Altarelli, 
hep-ph/0508053 and B. Kayser, hep-ph/0506165.
%
\bibitem{hirschetal}
M. Hirsch, H.V. Klapdor-Kleingrothaus, and S.G. Kovalenko, Phys. Lett. 
{\bf B398}, 311 (1997). 
%
\bibitem{grossman-haber1}
Y. Grossman and H.E. Haber, Phys. Rev. Lett. {\bf 78}, 3438 (1997).  
%
\bibitem{grossman-haber2}
Y. Grossman and H.E. Haber, Phys. Rev. D{\bf 59}, 093008 (1999); {\it 
ibid.} {\bf 63}, 075011 (2001).
%
\bibitem{chun}
E.J. Chun, Phys. Lett. {\bf B525}, 114 (2002).
%
\bibitem{davidson-king}
S. Davidson and S.F. King, Phys. Lett. {\bf B445}, 191 (1998).
%
\bibitem{bsystem}
For reviews, see, e.g. P.J. Franzini, Phys. Rep. {\bf 173}, 1 
(1989); H.R. Quinn, Eur. Phys. J. C {\bf 3}, 555 (1998).   
%
\bibitem{leptonflavor}
N. Arkani-Hamed, H-C. Cheng, J.L. Feng, and L.J. Hall, Phys. Rev. Lett. 
{\bf 77}, 1937 (1996).
%
\bibitem{staulsp}
A. Gould, B.T. Draine, R.W. Romani, and S. Nussinov, Phys. Lett. {\bf 
B238}, 337 (1990); G. Starkman, A. Gould, R. Esmailzadeh, and S. 
Dimopoulos, Phys. Rev. D{\bf 41}, 3594 (1990); T. Hemmick {\it et al.}, 
Phys. Rev. D{\bf 41}, 2074 (1990); P. Verkerk {\it et al.}, Phys. Rev. 
Lett. {\bf 68}, 1116 (1992).

\bibitem{like-sign-rp}
S. Kolb, M. Hirsch, H.V. Klapdor-Kleingrothaus, and O. Panella, Phys. Rev. 
D{\bf 64}, 115006 (2001).  
%
\bibitem{like-sign-norp}
K. Choi, K. Hwang, and W.Y. Song, Phys. Rev. Lett. {\bf 88}, 141801 (2002).
%
\bibitem{majoranasnu}
M. Hirsch, H.V. Klapdor-Kleingrothaus, S. Kolb, and S.G. Kovalenko, Phys. Rev. 
D{\bf 57}, 2020 (1998).

\bibitem{shaouly}
S. Bar-Shalom, G. Eilam, and A. Soni, Phys. Rev. Lett. {\bf 80}, 4629 (1998); Phys.
Rev. D{\bf 59}, 055012 (1999). 

\bibitem{amsb1}
L. Randall and R. Sundrum, Nucl. Phys. {\bf B557}, 79 (1999); G.F. 
Giudice, M.A. Luty, H. Murayama, and R. Rattazzi, J. High Energy Phys. 
{\bf 12}, 027 (1998).

\bibitem{amsbmod} A. Pomarol, R. Rattazzi, J. High Energy Phys. {\bf 05}, 013 (1999);
E. Katz, Y. Shadmi, Y. Shirman, J. High Energy Phys. {\bf 08}, 015 (1999);
R. Rattazzi, A. Strumia, J.D.~Wells, Nucl. Phys. {\bf B576}, 3 (2000);
Z.~Chacko, M. Luty, E. Pont\'{o}n, Y. Shadmi, Y. Shirman,
Phys. Rev. D{\bf 64}, 055009 (2001);
Z. Chacko, M.A. Luty, I. Maksymyk, E. Pont\'{o}n, J. High Energy Phys. 
{\bf 04}, 001 (2000);
I. Jack, D.R.T. Jones, Phys. Lett. {\bf B482}, 167 (2000);
N.~Arkani-Hamed, D.E.~Kaplan, H.~Murayama, Y.~Nomura,
J. High Energy Phys. {\bf 02}, 041 (2001);
M. Carena, K. Huitu, T. Kobayashi, Nucl. Phys.
{\bf B592}, 164 (2001).

\bibitem{amsb-hadron1}
J.L. Feng, T. Moroi, L. Randall, M. Strassler, and S. Su, Phys. Rev. Lett. 
{\bf 83}, 1731 (1999); T. Gherghetta, G.F. Giudice, and J.D. Wells, Nucl. 
Phys. {\bf B559}, 27 (1999); J.L. Feng and T. Moroi, Phys. Rev. D{\bf 61}, 
095004 (2000); S. Su, Nucl. Phys. {\bf B573}, 87 (2000); F. Paige and J. 
Wells, hep-ph/0001249; H. Baer, J.K. Mizukoshi, and X. Tata, Phys. Lett. 
{\bf B488}, 367 (2000). 

\bibitem{amsb-hadron2}
A. Datta, P. Konar and B. Mukhopadhyaya, Phys. Rev. Lett. {\bf 88}, 181802 (2002); 
A.J. Barr, C.G. Lester, M.A. Parker, B.C. Allanach, and P. Richardson, J. High 
Energy Phys. {\bf 03}, 045 (2003); A. Datta and K. Huitu, Phys. Rev. D{\bf 67}, 
115006 (2003).

\bibitem{amsb-linearee}
D.K. Ghosh, P. Roy, and S. Roy, J. High Energy Phys. {\bf 08}, 031 (2000); 
D.K. Ghosh, A. Kundu, P. Roy, and S. Roy, Phys. Rev. D{\bf 64}, 115001 
(2001); A. Datta and S. Maity, Phys. Lett. {\bf B513}, 130 (2001); 
M.A. D\'{\i}az, R.A. Lineros, and M.A. Rivera, Phys. Rev. D{\bf67}, 115004 
(2003). 

\bibitem{amsb-lineareg-gg}
D. Choudhury, D.K. Ghosh, and S. Roy, Nucl. Phys. {\bf B646}, 3 
(2002); D. Choudhury, B. Mukhopadhyaya, S. Rakshit, and A. Datta, J. High 
Energy Phys. {\bf 01}, 069 (2003). 

\bibitem{reviewamsb}
S. Roy, Mod. Phys. Lett. {\bf A19}, 83 (2004).

\bibitem{bilenky}
S. Bilenky, hep-ph/0509098.

\bibitem{cosmo-neut}
C. L. Bennett {\it et al.}, Astrophys. J. Suppl. {\bf 148}, 1 (2003); D. 
N. Spergel {\it et al.}, Astrophys. J. Suppl. {\bf 148}, 175 (2003); P. Crotty, J.
Lesgourgues, and S. Pastor, Phys. Rev. {\bf D69}, 123007 (2004); G.L. Fogli, E.
Lisi, A. Marrone, A. Melchiorri, A. Palazzo, P. Serra, and J. Silk, Phys. Rev. {\bf
D70}, 113003 (2004); U. Seljak {\it et al.}, Phys. Rev. {\bf D71}, 103515 (2005); S.
Hannestad, Phys. Rev. Lett. {\bf 95}, 221301 (2005). 

\bibitem{neutrinoless}
H.V. Klapdor-Kleingrothaus {\it et al.}, Mod. Phys. Lett., {\bf A16}, 2409 
(2001); H.V. Klapdor-Kleingrothaus, A. Dietz, I.V. Krivosheina, and O. 
Chkvorets, Nucl. Instrum. Meth. {\bf A 522}, 371 (2004); Phys. Lett. 
{\bf B586}, 198 (2004).  

\bibitem{sneu-double}
M. Hirsch, H.V. Klapdor-Kleingrothaus, and S.G. Kovalenko, Phys. Lett. 
{\bf B403}, 291 (1997).

\bibitem{egammacollider}
I.F. Ginzburg, G.L. Kotkin, V.G. Serbo, and V.I. Telnov, Nucl. Instrum. Methods 
{\bf 205}, 47 (1983); I.F. Ginzburg, G.L. Kotkin, S.L. Panfil, V.G. Serbo, 
and V.I. Telnov, {\it ibid.} {\bf 219}, 5 (1984).

\bibitem{egamma-tdr}
B. Badelek {\it et al.} [ECFA/DESY Photon Collider Working Group], Int. J. 
Mod. Phys. {\bf A19}, 5097 (2004).

\bibitem{chou-cuypers}
D. Choudhury and F. Cuypers, Nucl. Phys. {\bf B451}, 16 (1995).

\bibitem{berge}
S. Berge, M. Klasen, and Y. Umeda, Phys. Rev. D{\bf 63}, 035003 (2001).

\bibitem{egamma-prod}
V. Barger, T. Han, and J. Kelly, Phys. Lett. {\bf B419}, 233 (1998).

\bibitem{rpv-review}
For recent reviews on $R$-parity violation, see, e.g. R. Barbier {\it 
et al.}, Phys. Rep. {\bf 420}, 1 (2005); M. Chemtob, Prog. Part. Nucl. 
Phys. {\bf 54}, 71
(2005).

\bibitem{martin-vaughn}
S.P. Martin and M.T. Vaughn, Phys. Rev. D{\bf 50}, 2282 (1994).

\bibitem{potential} 
R. Arnowitt and P. Nath, Phys. Rev. D{\bf 46}, 3981 (1992); V. Barger, M.S. Berger,
and P. Ohmann, {\it ibid.}, {\bf 49}, 4908 (1994).

\bibitem{susy-qcd}
L.J. Hall, R. Rattazzi, and U. Sarid, Phys. Rev. D{\bf 50}, 7048 (1994); R.
Hempfling, {\it ibid.}, {\bf 49}, 6168 (1994); M. Carena, M. Olechowski, S.
Pokorski, and C. Wagner, Nucl. Phys. {\bf B426}, 269 (1994); D. Pierce, J. Bagger,
K.T. Matchev, and R. Zhang, {\it ibid.}, {\bf B491}, 3 (1997).

\bibitem{mass-diff}
H.C. Cheng, B.A. Dobrescu and, K.T. Matchev, Nucl. Phys. {\bf B543}, 47 (1999).

\bibitem{muong-2-1}
U. Chattopadhyay, D.K. Ghosh, S. Roy, Phys. Rev. D{\bf 62}, 115001 (2000).

\bibitem{PDG}
S. Eidelman {\it et al.}, Phys. Lett. {\bf B592}, 1 (2004).

\bibitem{haber-carena}
M. Carena and H.E. Haber, Prog. Part. Nucl. Phys. {\bf 50}, 63 (2003).

\bibitem{muong-2-2}
J.L. Feng, K.T. Matchev, Phys. Rev. Lett. {\bf 86}, 3480 (2001); K. Choi, K. Hwang,
S.K. Kang, K.Y. Lee, W.Y. Song, Phys. Rev. D{\bf 64}, 055001 (2001).

\bibitem{muong-2-3}  
U. Chattopadhyay, P. Nath, Phys. Rev. Lett. {\bf 86}, 5854 (2001); H. Baer, C.
Balazs, J. Ferrandis, X. Tata, Phys. Rev. D{\bf 64}, 035004 (2001); K. Enqvist, E.
Gabrielli, K. Huitu, Phys. Lett. {\bf B512}, 107 (2001).

\bibitem{lightbylight}
M. Knecht, A. Nyffeler, Phys. Rev. D{\bf 65}, 073034 (2002); M. Knecht, A.
Nyffeler, M. Perrottet, E. de Rafael, Phys. Rev. Lett. {\bf 88}, 071802 (2002); M.
Hayakawa, T. Kinoshita, hep-ph/0112102; I. Blokland, A. Czarnecki, K. Melnikov,
Phys. Rev. Lett. {\bf 88}, 071803 (2002); J. Bijnens, E. Pallante, J. Prades, Nucl.
Phys. {\bf B626}, 410 (2002); M. Ramsey-Musolf and M.B. Wise, Phys. Rev. Lett. {\bf
89}, 041601 (2002).

\bibitem{E821}
Muon g -- 2 Collaboration, G.W. Bennett, {\it et al.}, Phys. Rev. Lett. {\bf 89},
101804 (2002).

\bibitem{vacuumstability}
A. Datta, A. Kundu, A. Samanta, Phys. Rev. D{\bf 64}, 095016 (2001); E. Gabrielli,
K. Huitu, S. Roy, Phys. Rev. D{\bf 65}, 075005 (2002).

\bibitem{universe}
A. Riotto, E. Roulet, Phys. Lett. {\bf B 377}, 60 (1996); A. Kusenko, P. Langacker,
G. Segre, Phys. Rev. D{\bf 54}, 5824 (1996).

\bibitem{rpv-bounds}
V. Barger, G.F. Giudice, and T. Han, Phys. Rev. D{\bf 40}, 2987 (1989); 
B.C. Allanach, A. Dedes, and H.K. Dreiner, Phys. Rev. D{\bf 60}, 075014 (1999); 
F. Ledroit and G. Sajot, Report No. GDR-S-008 (ISN, Grenoble, 1998). This can be
obtained at 
http://qcd.th.u-psud.fr/GDR$\_$SUSY/GDR$\_$SUSY$\_$PUBLIC/entete$\_$note$\_$publique

\bibitem{pilaftsis}
S. Bray, J.S. Lee, and A. Pilaftsis, Phys. Lett. {\bf B628}, 250 (2005).

\end{thebibliography}
\end{document}